\documentclass[aps,prl,preprint]{revtex4-2}

\usepackage{amsmath}
\usepackage{graphicx}
\usepackage{float}

\begin{document}

\title{First Direct Observations of Gear-Changing In A Collider}
\author{E.~Nissen}
\email{nissen@jlab.org}
\affiliation{Thomas Jefferson National Accelerator Facility, Newport News, VA, USA}
\author{A.~K\"{a}llberg}
\email{anders.kallberg@fysik.su.se}
\author{A.~Simonsson}
\email{ansgar.simonsson@fysik.su.se}
\affiliation{Stockholm University, Stockholm, Sweden}

\begin{abstract}
In this work we perform the first ever demonstration of gear-changing in a real world collider. Gear-changing refers to a collision scheme where each ring of a collider stores a different harmonic number of bunches. These bunches are kept synchronized using different velocities. Such a system has been theorized, but has now been demonstrated using the Double ElectroStatic Ion Ring ExpEriment (DESIREE) in Stockholm Sweden. The experiment was able to demonstrate a gear-changing system, with both four on three bunches and five on four bunches. We determined a measurable parameter that shows a gear-changing system out to $37500$ turns of the slow beam. We also developed new insights into how to control this type of system, opening up new possibilities for research.
\end{abstract}

\maketitle

\section{Introduction}
\label{introduction}
\begin{figure}[H]
\centering
\includegraphics[width=0.5\linewidth]{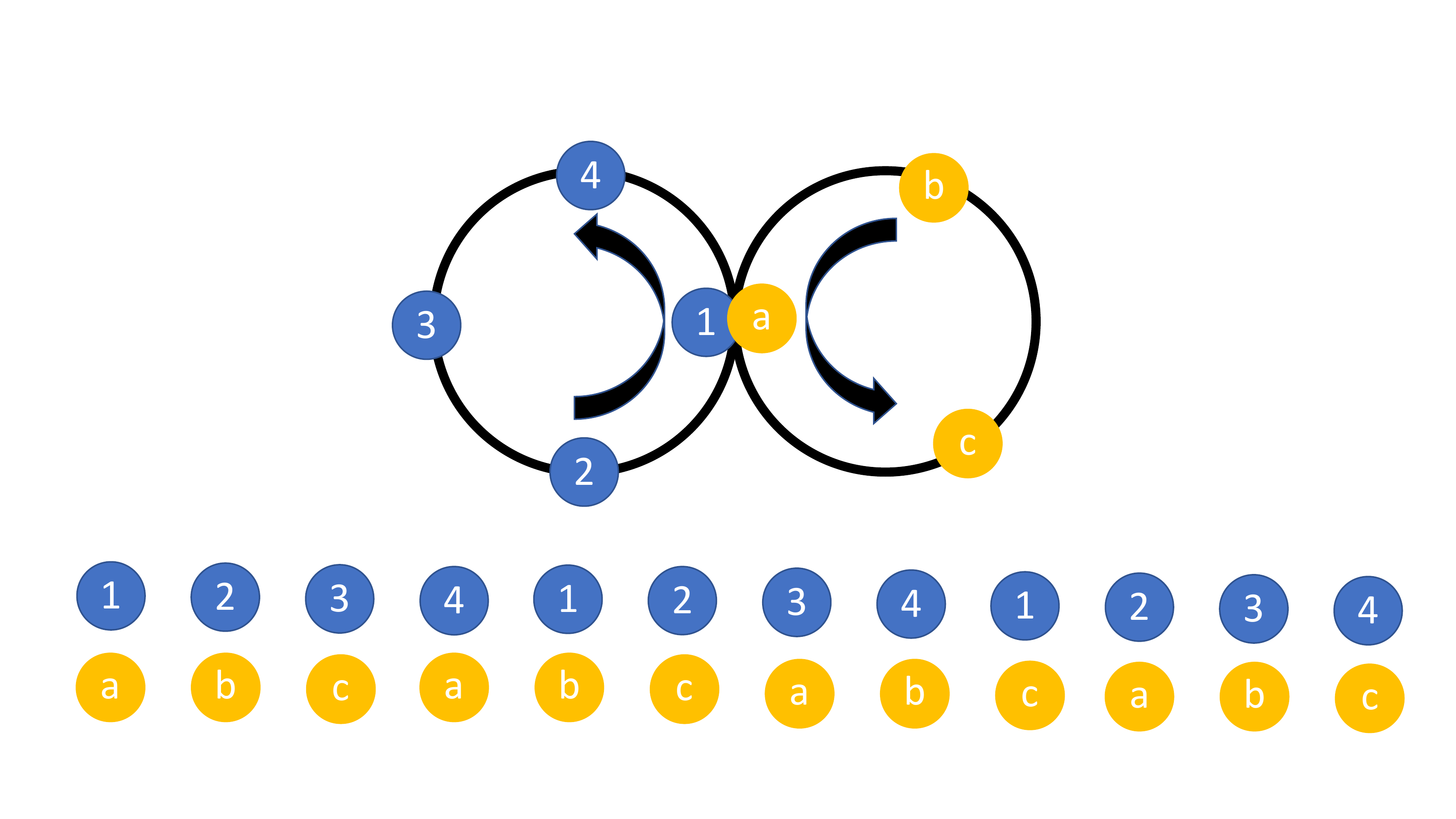}
\caption{This is a schematic view of a four on three type of gear-changing system with head-on collisions. As can be seen in each bunch will interact with each other bunch in a repeating pattern.} 
\label{fig_GearChange}
\end{figure}

Gear-changing refers to a system where two collider rings have different harmonic numbers and different bunch numbers while still having all bunches in the two rings collide with each other. For example, a collider could have four bunches in one ring and three bunches moving at 4/3 the velocity in the other ring. The synchronization could also be achieved with an appropriate pathlength difference between the two rings. A system utilizing gear-changing allows for a wide range of energies and particle species without changing the size of the rings. It can also reduce systematic errors in some experiments, i.e. where spin polarization is important. A schematic of gear-changing is shown in Fig \ref{fig_GearChange}.

Gear-changing also has some drawbacks. Since each bunch interacts with each other bunch, each bunch now has to come into equilibrium all the other bunches in the machine. In systems like Super KEK-B or the Jefferson Lab EIC proposal the number of bunches is in the thousands, and can be approximated by a linac-ring type system \cite{Nissen:IPAC2017-THPAB082} \cite{Nissen:2019szv}. This type of system leads to stability issues and cross resonances \cite{HIRATA1990156} \cite{PhysRevSTAB.17.041001}.

There is also a problem caused by each beam's abort gap which means that when gear-changing reaches these gaps the beams will not receive any type of beam-beam effect. A theoretical treatment was performed for possible use in RHIC to reduce systematic errors \cite{PhysRevSTAB.17.041001}. An experiment to determine the effects of abort gaps was performed using an electron lens to simulate the interactions \cite{Luo:2014dca}. Demonstrating a full gear-changing system is possible in RHIC, though it is invasive and has not been tried. Another machine is able to demonstrate gear-changing at lower energies,  the Double ElectroStatic Ion Ring ExpEriment (DESIREE), in Sweden \cite{doi:10.1063/1.3602928}. 

DESIREE is made up of a pair of electrostatic storage rings that can merge up to 25 keV in one ring and up to 100 keV in the other ring beams of a variety of mostly singly charged ions \cite{PhysRevA.102.012823}. These beams merge with each other while moving in the same direction. Typically velocities are matched to measure mutual neutralization reactions like those that would be found in the interstellar medium. The construction allows the use of a wide range of relative velocities, and since the velocities are different in our experiment, we will have collisions in a moving reference frame. This experiment demonstrated these gear-changing systems, using $14.3$ keV nitrogen and $6.97$ keV carbon test beams. 

\section{Experimental Setup}
\label{expsetup}

\begin{figure}[H]
\centering
\includegraphics[width=0.5\linewidth]{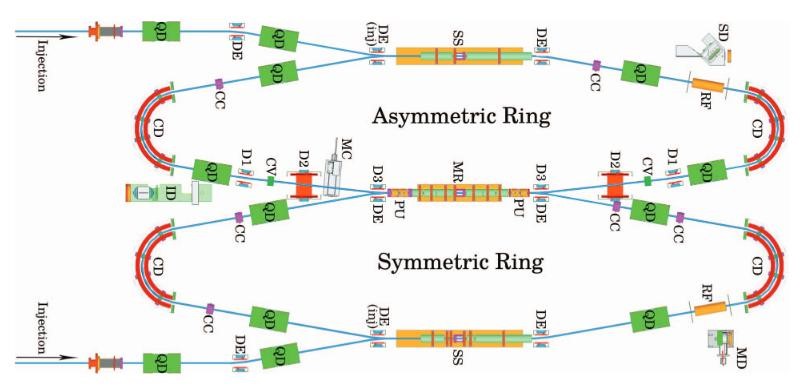}
\caption{This is a schematic diagram of DESIREE. The collisions occur in the merger region (MR) and are measured using the pickups (PU) before and after the merger region. PU1 is on the right side of the MR (upstream) and PU2 is on the left side MR in this figure (downstream).} 
\label{fig_DESIREE}
\end{figure}

DESIREE consists of two electrostatic ion-beam storage rings with a 1 m long merging section, where the two stored ion beams overlap. The storage rings are enclosed in a single cryogenically cooled vacuum chamber, operating at 13 K with a residual gas density of about $10^4$ H$_2$ molecules per cm$^3$. In order to merge two ion beams with different kinetic energies, one of the rings has been equipped with four additional deflectors to compensate for the different bending angles in the common bends. With these deflectors beams with different kinetic energies up to a factor of 20 can be merged. To make space for the extra deflectors, the four quadrupole lenses in this ring are unevenly distributed, it is referred to as the asymmetric ring. It has a circumference of 8.71 m $\pm$2 cm depending on settings. The quadrupoles are evenly distributed around the other ring, the symmetric ring, which has a circumference of 8.68 m. The ion beam with the highest kinetic energy is stored in the asymmetric ring. A diagram of DESIREE is shown in Fig \ref{fig_DESIREE}.

Two fast choppers bunch the beams to get one or several bunches with a square time-current bunch shape, with a 1.2 $\mu$s pulse length. To get the required revolution frequencies, corrections were done in two steps. First the ion source platform voltages were adjusted, including rescaling of all bends and quadrupoles. Next the voltages on the drift tubes were changed to locally vary the velocities of both beams and thus also the revolution frequencies. The frequencies were measured with a Fourier transform of the pick-up signal without RF.

RF is not normally used in DESIREE, but here it was applied on two 20 cm long drift tubes which otherwise are used as Schottky detectors. This was used to maintain the bunch structure over long periods of time \cite{Nissen:DODGE0}. In this experiment we used a sinusoidal waveform with amplitudes of $1$-$4$ V. To make the bunches of the two rings overlap in the common straight section, the timing of the choppers and the RF phases were adjusted.

\begin{figure}[H]
\centering
\includegraphics[width=0.5\linewidth]{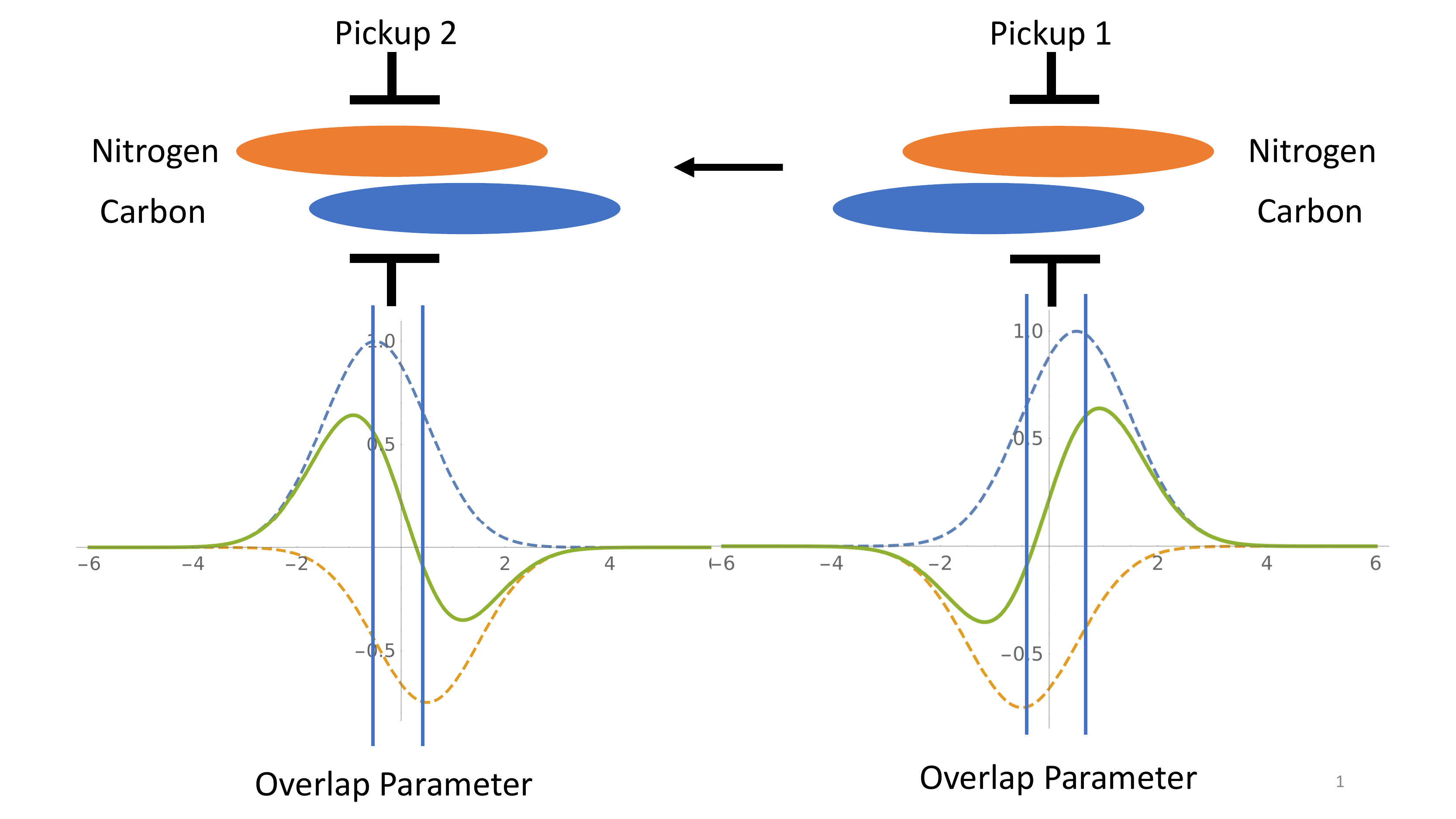}
\caption{This diagram shows how the bunches collide while moving through the pickups. The dashed lines show the raw signals for the carbon and nitrogen bunches, with the solid line showing the pickup signal. The distance between these peaks is the overlap parameter.} 
\label{fig_overlap}
\end{figure}

The core of this experiment is a four on three gear-changing system. Carbon was chosen for the slow (four bunch) and nitrogen was chosen for the fast (three bunch) beams. Since these bunches are oppositely charged, their oscilloscope signals will cancel each other in the pickups. The pickup amplification is -500, so a bunch of N$^{1+}$ gives a negative signal, while a C$^{1-}$ bunch gives a positive signal. Because the nitrogen beam is 33\% faster than the carbon beam there will be an overlap pattern in pickup 1 (PU1) and a mirror of this overlap pattern in pickup 2 (PU2). This occurs because the carbon bunch starts out ahead of the nitrogen bunch, but since the nitrogen bunch is faster it passes through the carbon bunch and exits before the carbon. The bunches start out with a square longitudinal profile which evolves into a Gaussian. The exact pattern will evolve into something similar to what is shown in Fig \ref{fig_overlap}.

If we assume that the bunches are perfectly timed such that they cross at the center of the merger region then the profiles in PU1 and PU2 will be symmetric. The distance between the center of the carbon and nitrogen beams is referred to as the overlap parameter. The demonstration of the gear-changing system has two parts. First, we will use an RF bucket filling scheme where the last bunches of each ring will be left off, this will then have three carbon bunches with the fourth bucket empty, and two nitrogen bunches with the third bucket empty. This will result in a repeating pattern every three carbon turns which will quickly show that gear-changing is occurring. It will also allow us to check how the bunches are evolving, and how that affects our ability to measure the distance between the beams. We call this the missing bunch experiment Once everything is measured out properly we will measure the beams with full buckets. This is the second part, the full bucket experiment.

The beam-beam tune shift for a machine like this is extremely small, due to the low charge to mass ratio and small particle number. There will be a small boost from having the beams moving in the same direction, however we used an algorithm \cite{Nissen:Gear} to perform some simple simulations. A 1 mm offset in one plane of a carbon bunch would give a 1.2 nm  offset to a nitrogen bunch in that plane after 1000 carbon turns, this is not measurable.

\section{Experimental Measurements}
\label{expmeas}
\begin{figure}[H]
\centering
\includegraphics[width=0.5\linewidth]{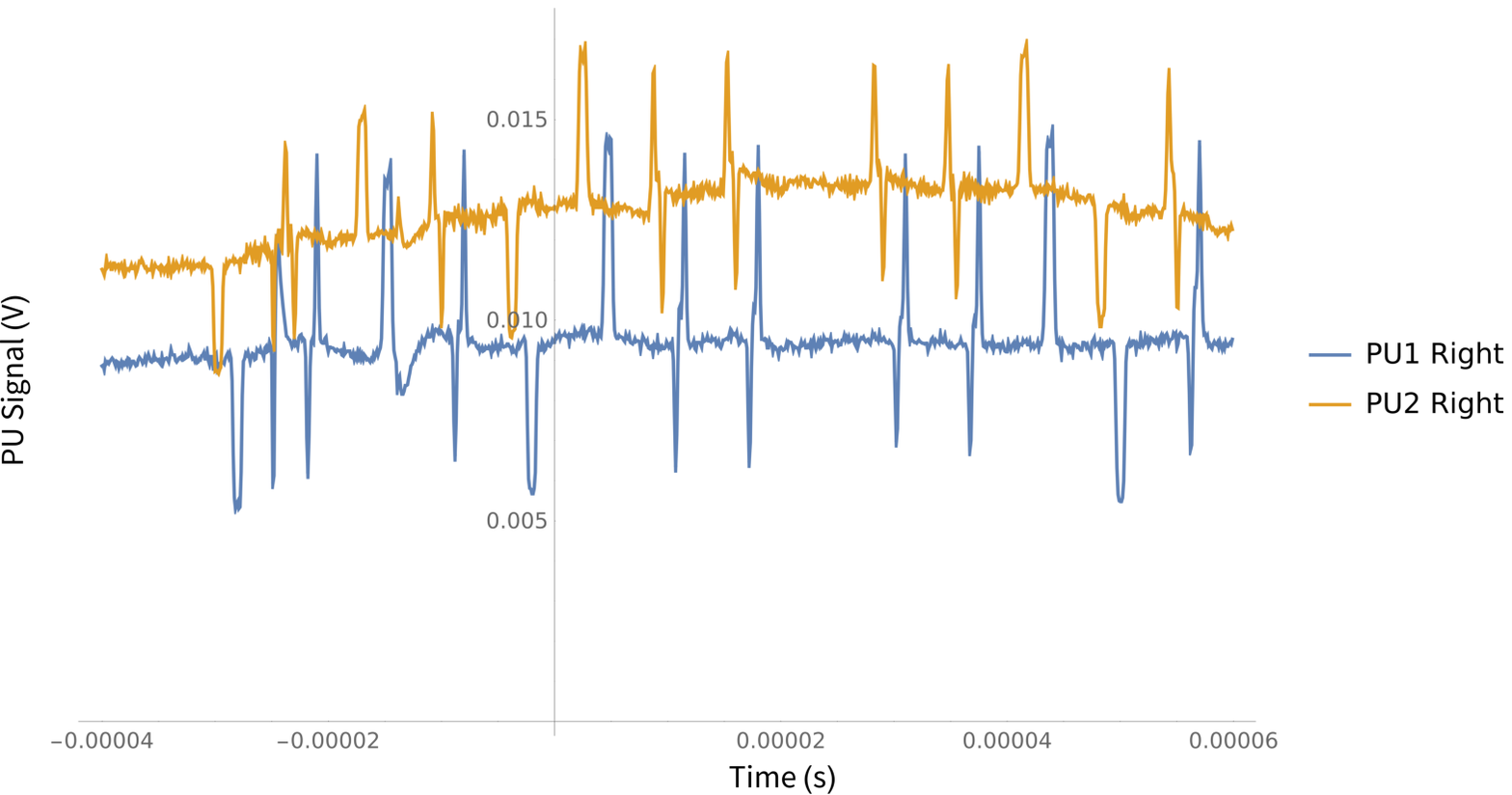}
\includegraphics[width=0.5\linewidth]{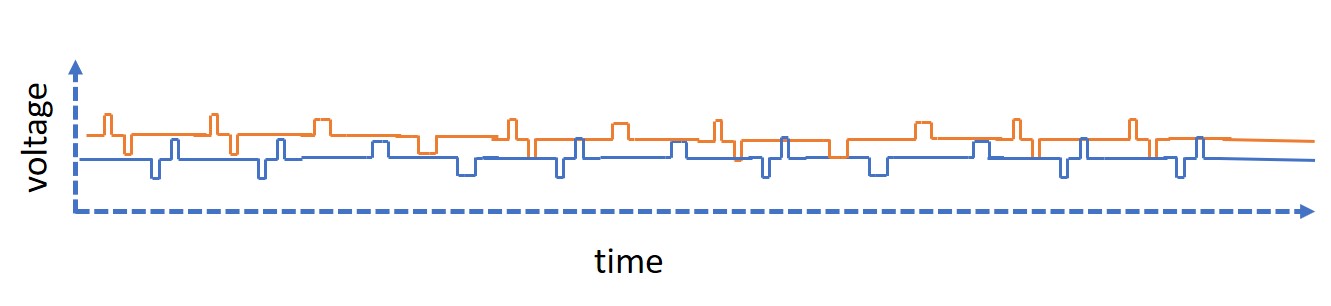}
\caption{A plot of the missing bunch pattern as read by the oscilloscope, using signals from the right side of each Pickup, the left signals are omitted in this plot for clarity. The lower plot shows a prediction of this pattern.} 
\label{fig_missing_bunch}
\end{figure}

The experiment was performed in May of 2020, via a remote collaboration because of COVID-19 related travel restrictions. Due to the small size of the offsets in the pickup signal we used an oscilloscope that would take snapshots of the beam at different numbers of turns. We can see an example of the repeating pattern in Fig \ref{fig_missing_bunch}. The symmetric pattern in the collisions is visible, as is the shape of the bunches without collisions, as well as the delay between signals from the pickups before and after the MR. These plots show the pickup signal averaged over 32 test runs.

In order to measure the collision patterns we analyzed the data by fitting the oscilloscope signals to double Gaussians with a linear baseline. The fit function we use is,

\begin{equation}
M = f \times e^{\frac{-(x-a)^2}{2*b^2}}-g \times e^{\frac{-(x-c)^2}{2*d^2}}+E+x\times h.
\end{equation}

Where $a$, $b$, $c$, $d$, $E$, $f$ and $g$ are fitting parameters. We zoomed in on the relevant collisions for each oscilloscope channel, and used a fitting function with the initial positions and voltages of the carbon and nitrogen bunches. This is the unconstrained fit. We can then plot the differences between the pickup positions of the peaks of the carbon and nitrogen going in and coming out of the merger region. For perfectly adjusted beams these distances should be the same on either side of the merger region. 

Since the bunches are not perfect Gaussians, the fits may not necessarily show the reality of the system. One method of countering this is to constrain the possible fit functions. We do this using the magnitude and $\sigma$ that we obtain from the uncollided bunches in the missing bunch system. We then constrain the fitting function to stay within 10\% of the $\sigma$ values and 25\% of the magnitude values. This is the constrained fit.

\section{Missing Bunch Experiment}
\label{missmeas}

We start benchmarking the system by zooming in on the uncollided bunches. This will give us an idea of the spacing of our bunches, as well as a measurement of their bunch lengths and the magnitudes of their signals. A variety of data sets were taken with different sweep lengths, a full turn for the carbon beam takes 26 $\mu$s. In order to get sufficient coverage, for this initial measurement we used a 50 $\mu$s sweep. The error bars shown are the step sizes of the oscilloscope sweeps.

\begin{figure}[H]
\centering
\includegraphics[width=0.5\linewidth]{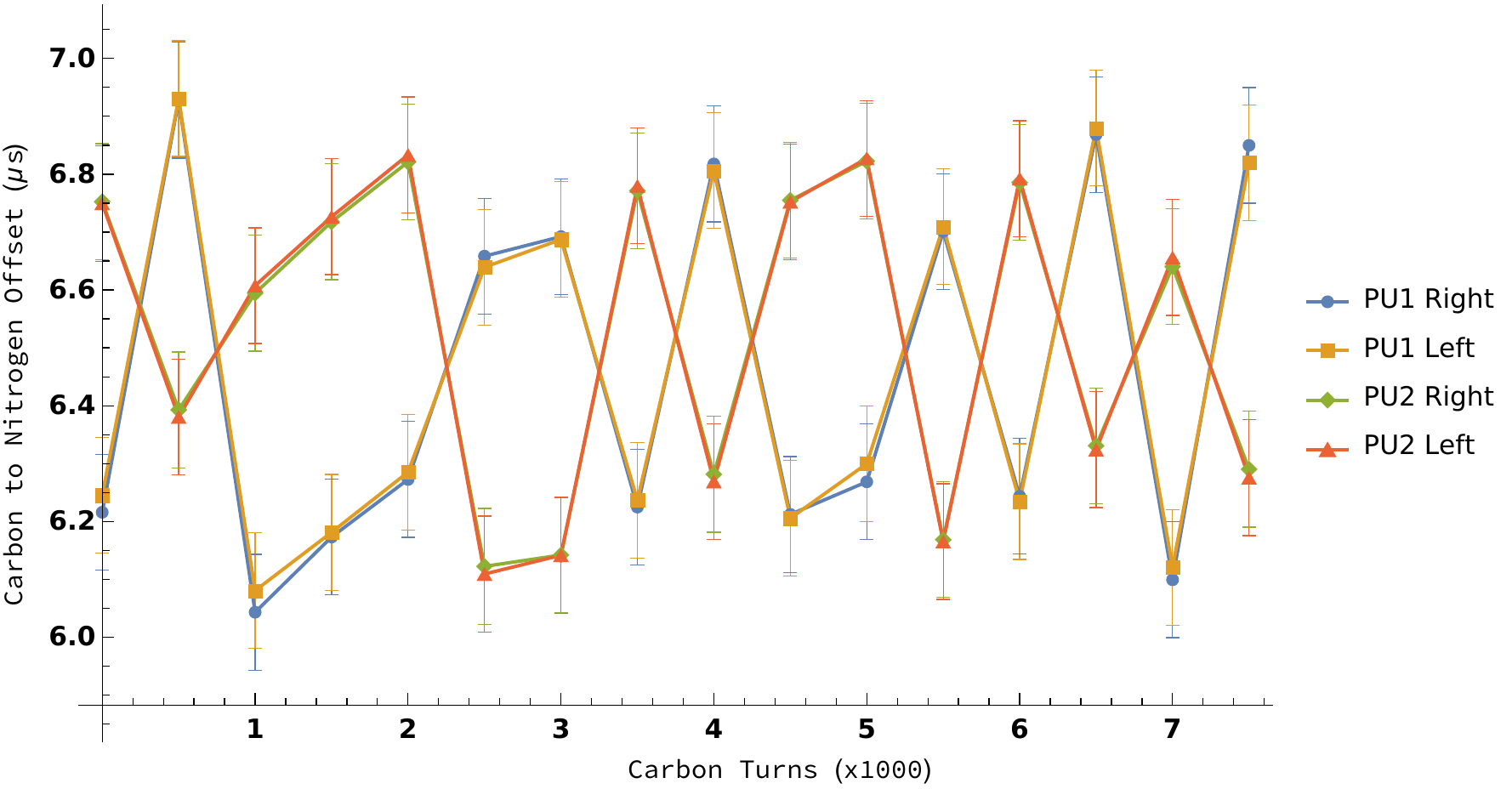}
\caption{This measurement takes the paired uncollided bunches in the missing bunch experiment up to 7500 carbon turns and measures the time difference between the carbon and nitrogen bunches. } 
\label{fig_7500_offset}
\end{figure}

By measuring the difference between the centers of the carbon and nitrogen bunches in the uncollided beam, we can check the timing of the system. A plot of this is shown in Fig \ref{fig_7500_offset}. The traces switch sides due to the fact that we are sampling different portions of the repeating pattern, some show the carbon going through the pickup first, others second. The bunch spacing between carbon bunches is 6.5 $\mu$s, so the time difference between where the nitrogen bunches fall will be plus or minus the expected overlap parameter that we wish to measure in the gear-changing experiment. 

When looking at the average difference between the bunches we can see that the overlap parameter comes out to 0.2718 $\mu$s, with a $\sigma$ of 6 ns. The distance between the centers of the pickups is 77 cm. The expected overlap parameter based on the energies of the particles and the 25 V on the electrode in the MR is 0.2803 $\mu$s, and the energies of the ions are accurate to within 50 V. This spread in energy does provide enough of an uncertainty that, combined with the 50 ns stepsize of the oscilloscope measurements, we feel that the measured overlap parameter is consistent with predictions.

\begin{figure}[H]
\centering
\includegraphics[width=0.49\linewidth]{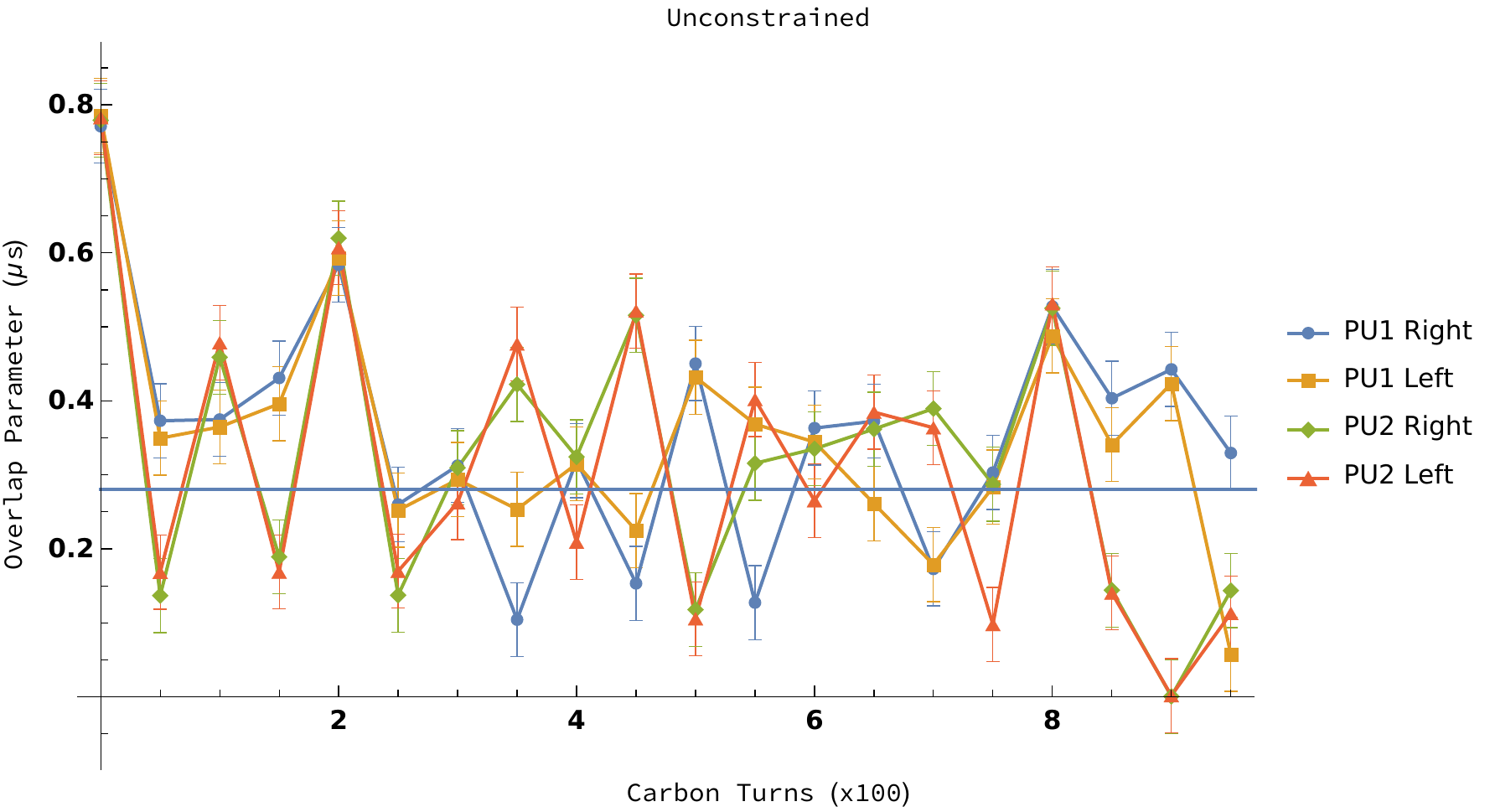}
\includegraphics[width=0.49\linewidth]{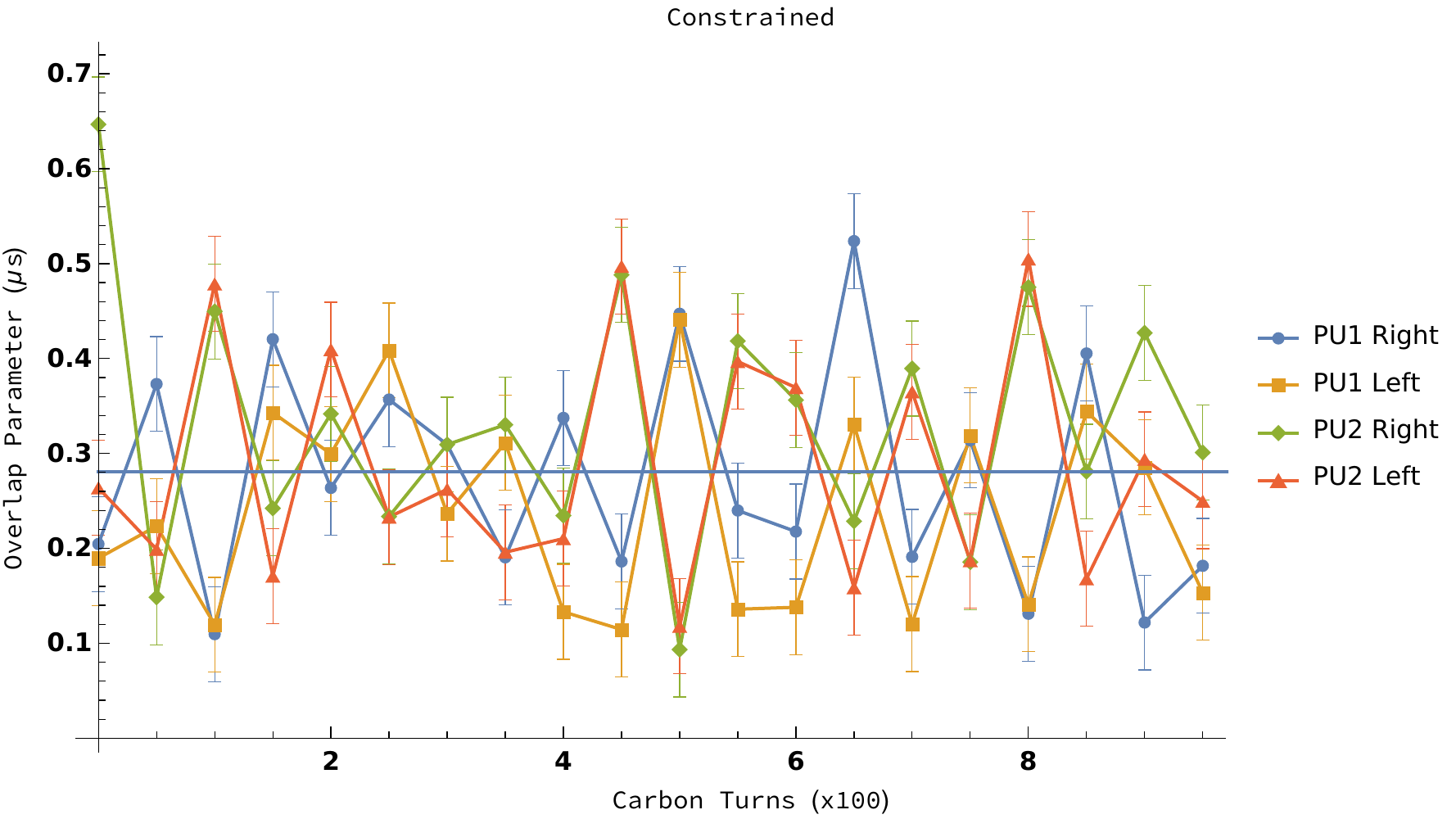}
\caption{These plots show the evolution of the overlap parameter using first the unconstrained and then the constrained fitting method for the bunches out to 950 carbon turns. The horizontal line shows the predicted overlap parameter.} 
\label{fig_950_missing}
\end{figure}

The next step is to look at the colliding bunches and measure their overlap parameters. Examples of the results from the different fitting algorithms for colliding bunches in the missing bunch experiment are shown in Fig \ref{fig_950_missing}. These show the overlap parameters oscillating around their expected values

\section{Full Bucket Experiment}
\label{FGMea}

\begin{figure}[H]
\centering
\includegraphics[width=0.49\linewidth]{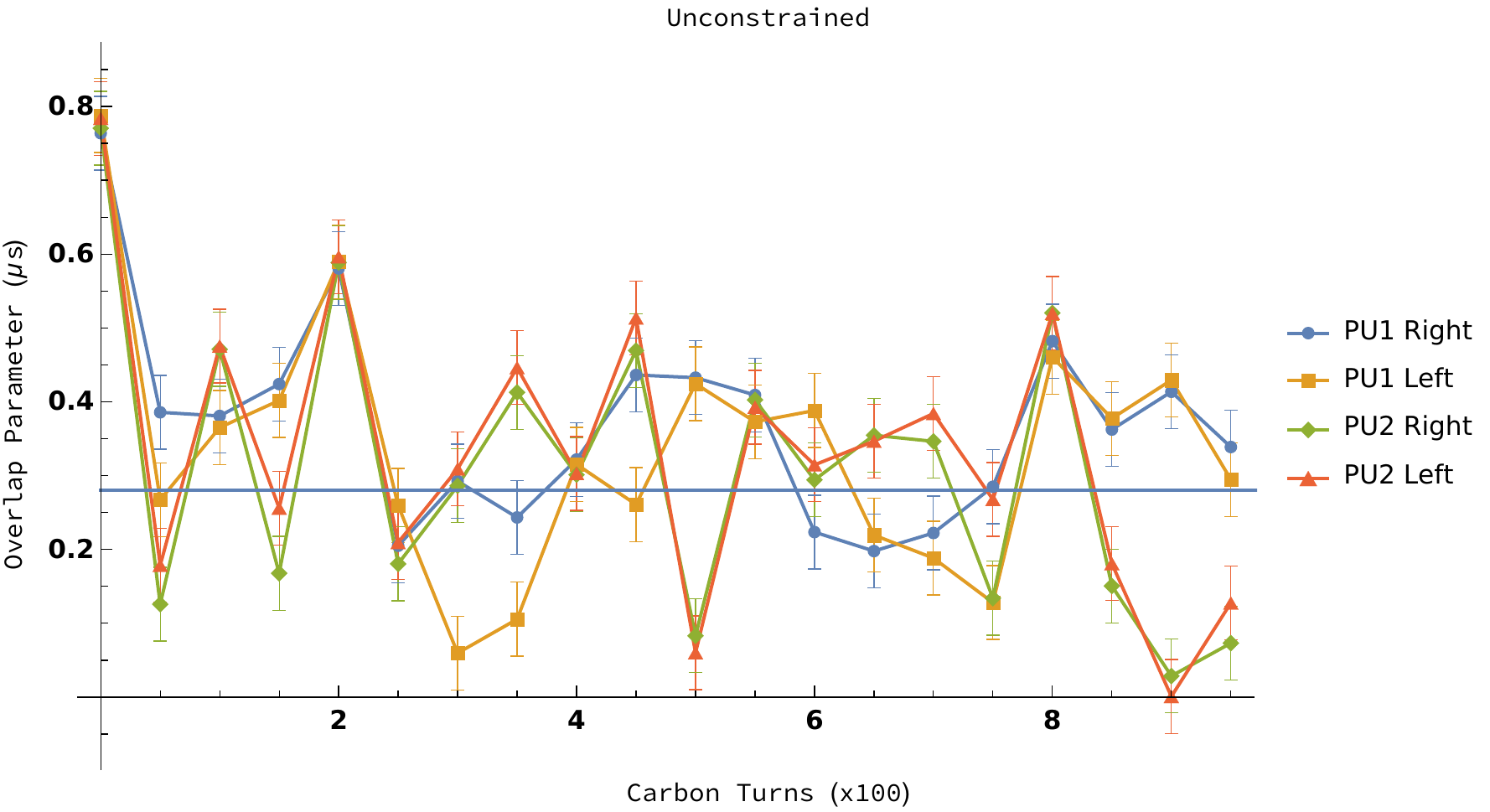}
\includegraphics[width=0.49\linewidth]{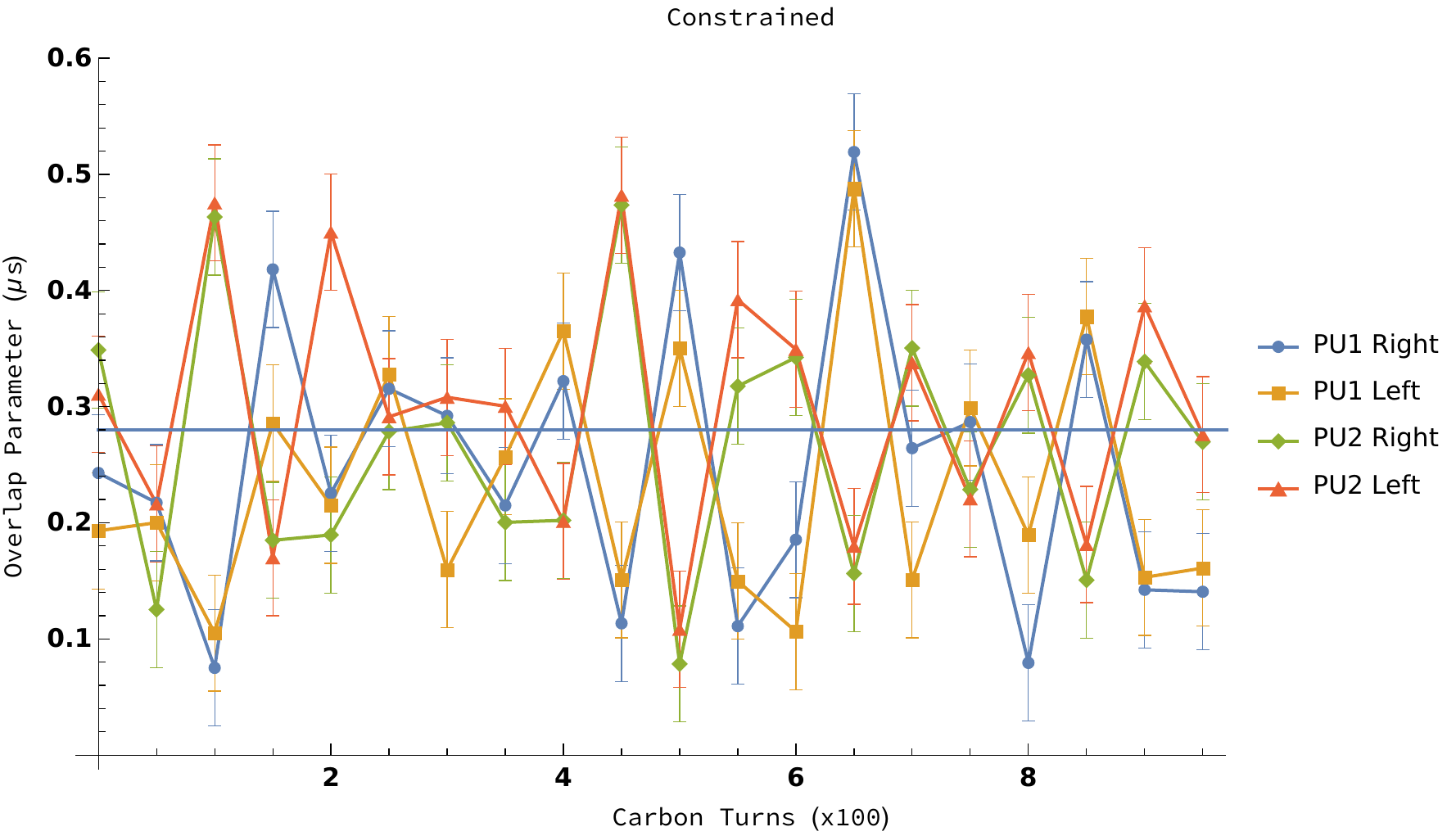}
\caption{These plots show the evolution of the overlap parameter using first the unconstrained and then the constrained fitting method for the bunches out to 950 carbon turns in the missing bunch experiment. The horizontal line shows the predicted overlap parameter.} 
\label{fig_950_full}
\end{figure}

\begin{figure}[H]
\centering
\includegraphics[width=0.49\linewidth]{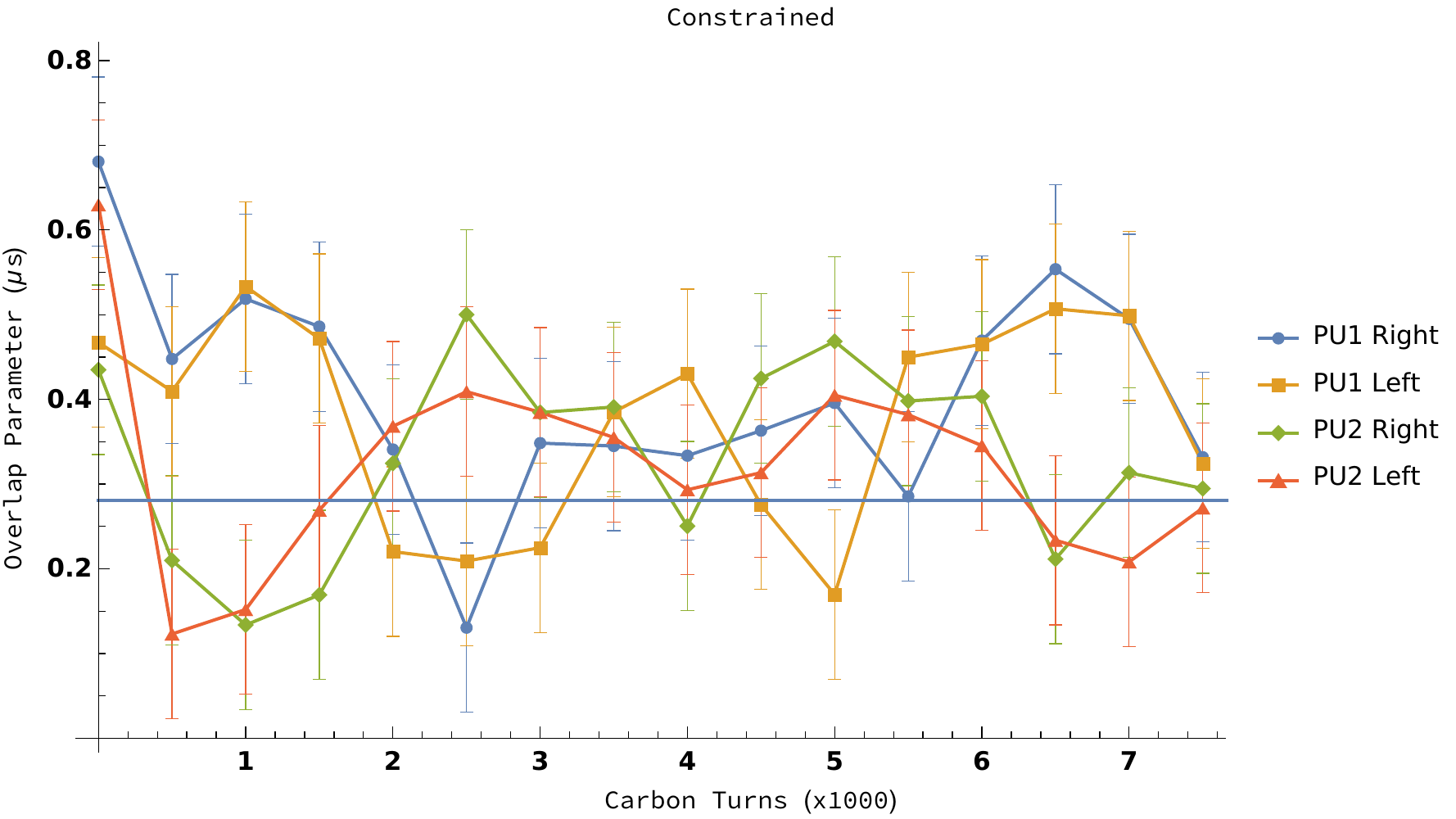}
\includegraphics[width=0.49\linewidth]{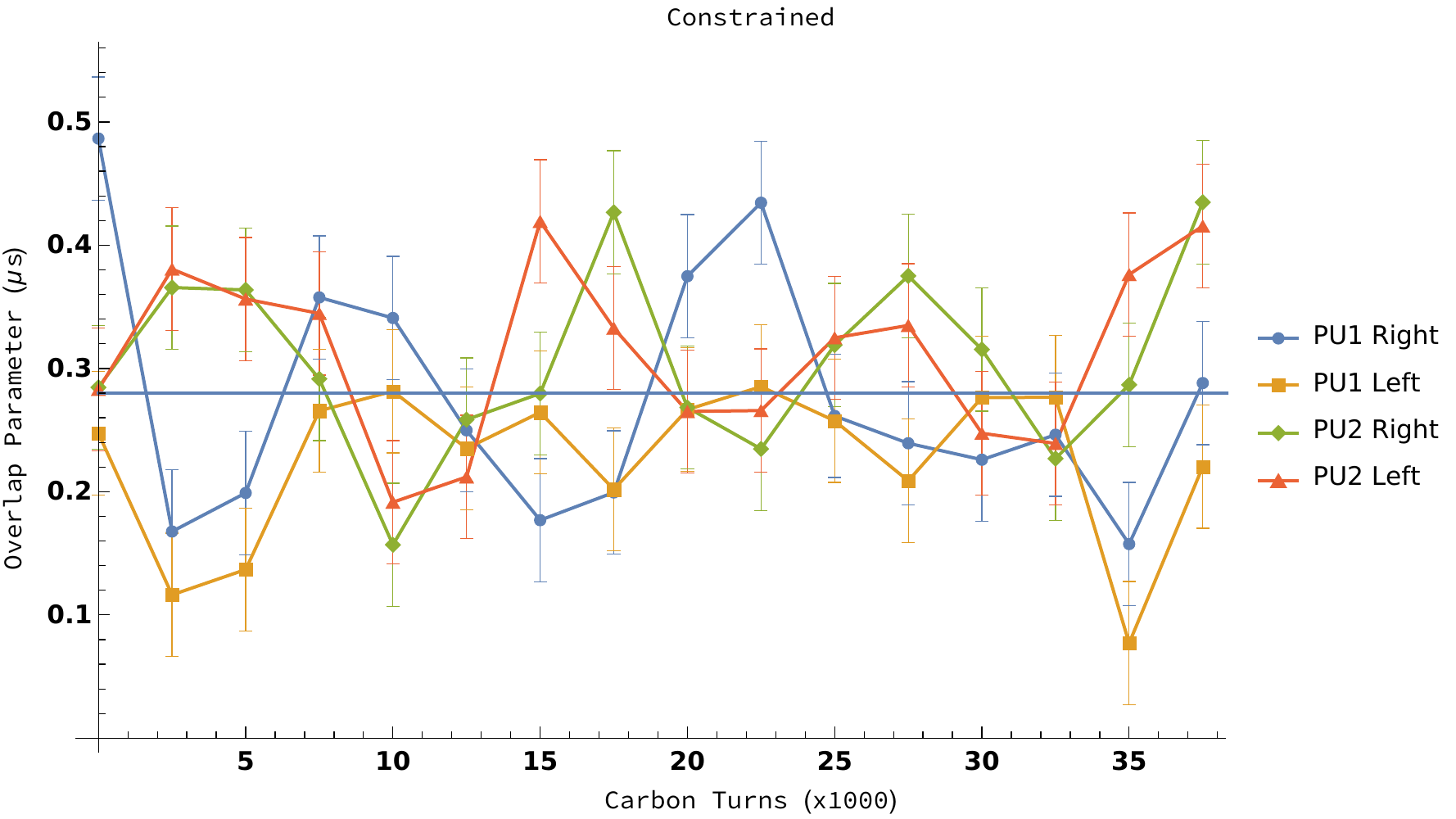}
\caption{These plots show the evolution of the overlap parameter using the constrained fitting method for the bunches out to 7500  carbon turns in the first plot, and 37500 carbon turns in the second. The horizontal line shows the predicted overlap parameter.} 
\label{fig_7500_full}
\end{figure}

Now that the collisions have been benchmarked and the overlap parameter is understood we measured the four on three gear-changing system while filling all buckets. Using a similar 50 $\mu$s sweep, we can use the constrained results to perform these measurements. These are shown in Figs \ref{fig_950_full}, and \ref{fig_7500_full}.

We synchronized the system in this experiment by trying to keep the patterns in the pickups as symmetric as possible. We adjusted the phase difference between the two RF cavities, which only move in $1^\circ$ increments. This gives a time resolution of 18 ns. Furthermore, both cavities are being run at the same frequency despite the two rings having slightly different pathlengths. The voltages on the electrodes in the MR should help compensate for this. The energies of the two beams are accurate to 50 V. These combinations of factors could lead to some of the oscillations seen in the overlap parameter in this experiment. Now that we have shown that it is possible to create a gear-changing system, future experiments can systematically work to reduce these errors.

\section{Five on Four Experiment}
\begin{figure}[H]
\centering
\includegraphics[width=0.5\linewidth]{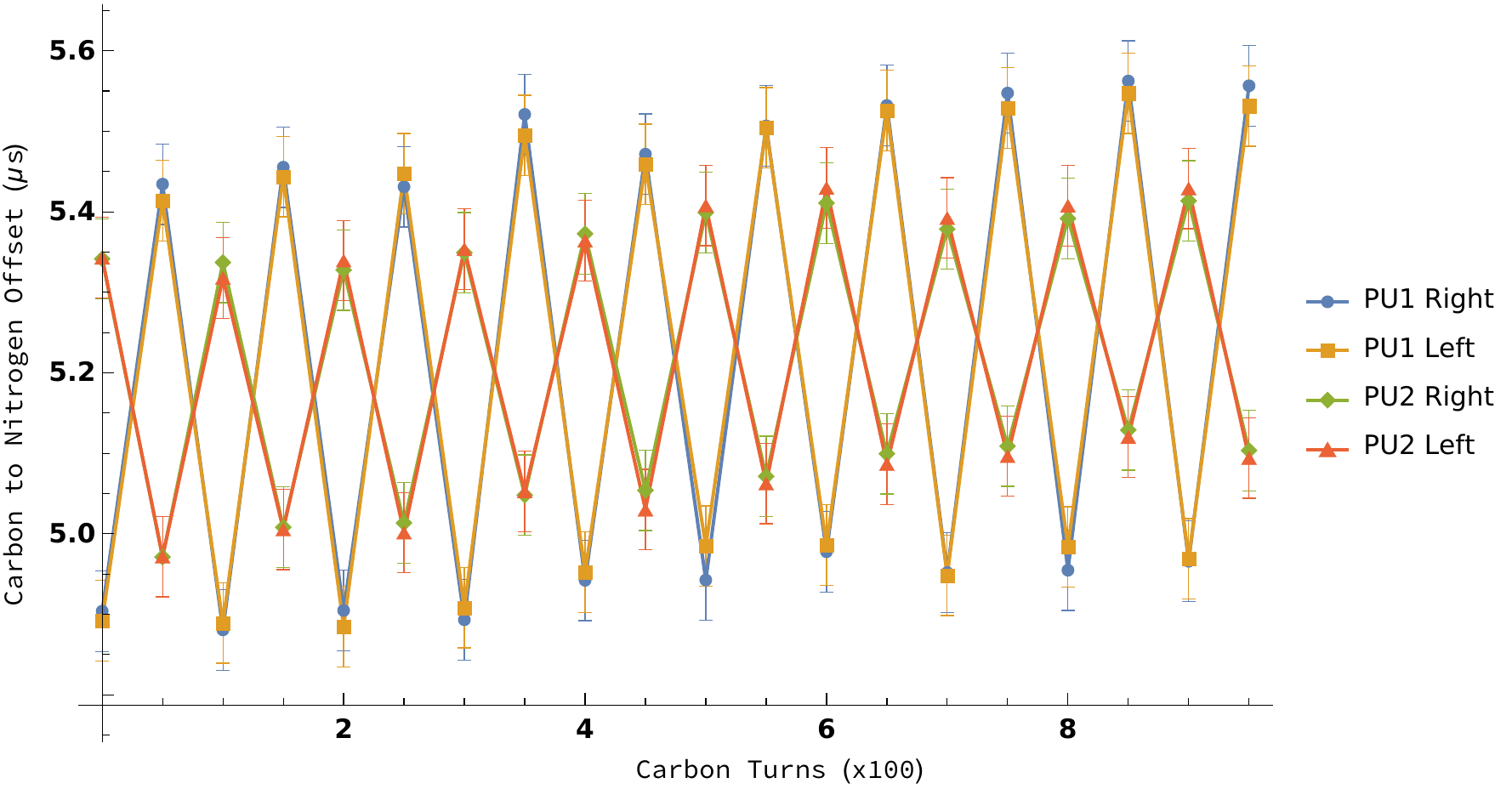}
\caption{This measurement takes the paired uncollided bunches in the missing bunch experiment up to 950 C-turns and measures the time difference between the carbon and nitrogen bunches.} 
\label{fig_950_54_offset}
\end{figure}

Another iteration of these measurements used a five on four gear-changing setup. We maintained the same 6.97 keV carbon energy and added another bunch. We decreased the nitrogen energy to 12.515 keV to match the velocity to 125\% of the carbon velocity and also added another bunch. This leads to a smaller overlap parameter since the velocity difference is smaller. We measured the uncollided bunches in Fig \ref{fig_950_54_offset}, and we have a commensurate difference of .22 $\mu$s with a $\sigma$ of 7 ns. The predicted overlap parameter is 0.215 $\mu$s, with the same MR distance and an electrode voltage of 70 V. Again these values are in agreement.

\begin{figure}[H]
\centering
\includegraphics[width=0.49\linewidth]{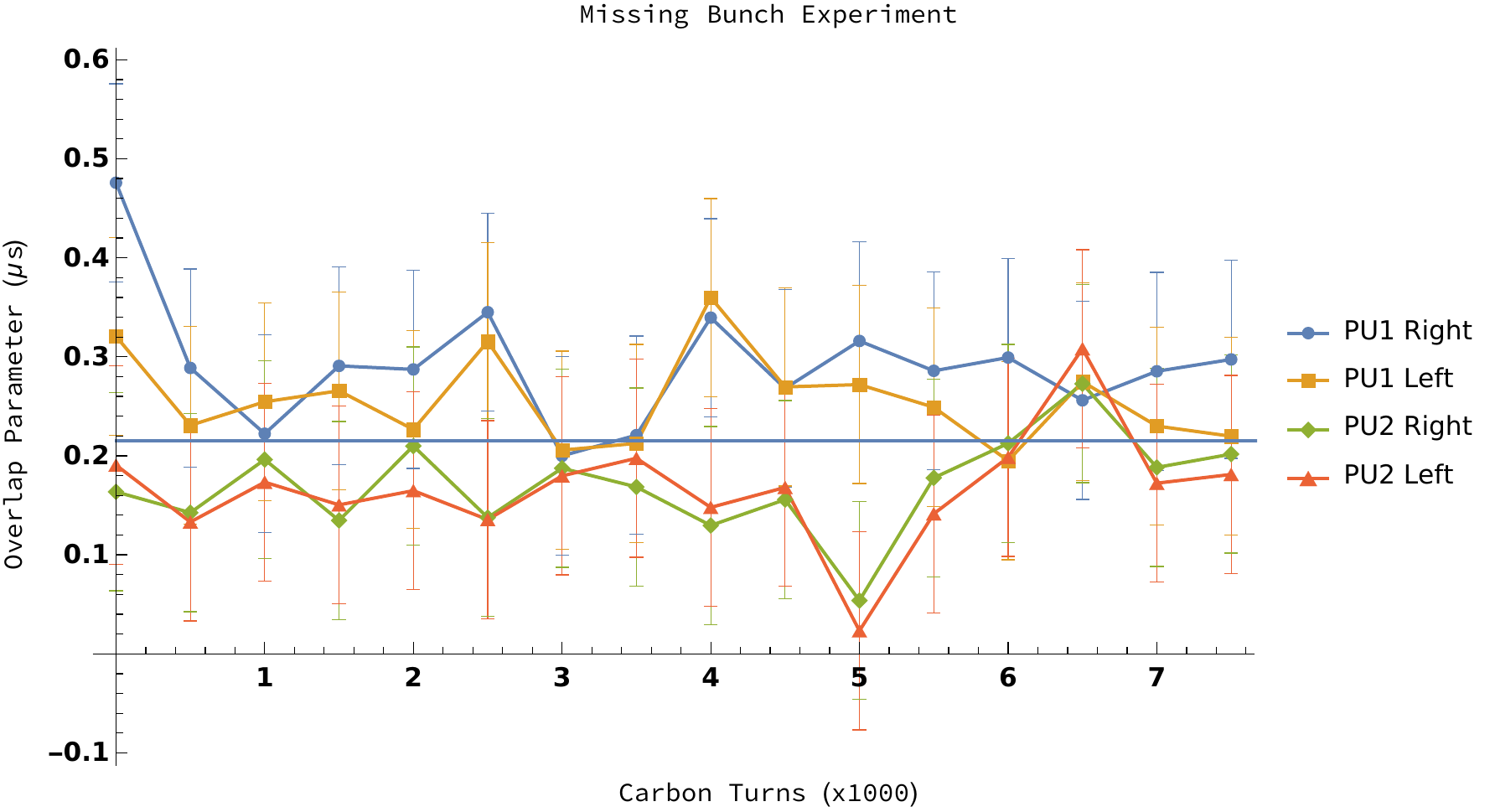}
\includegraphics[width=0.49\linewidth]{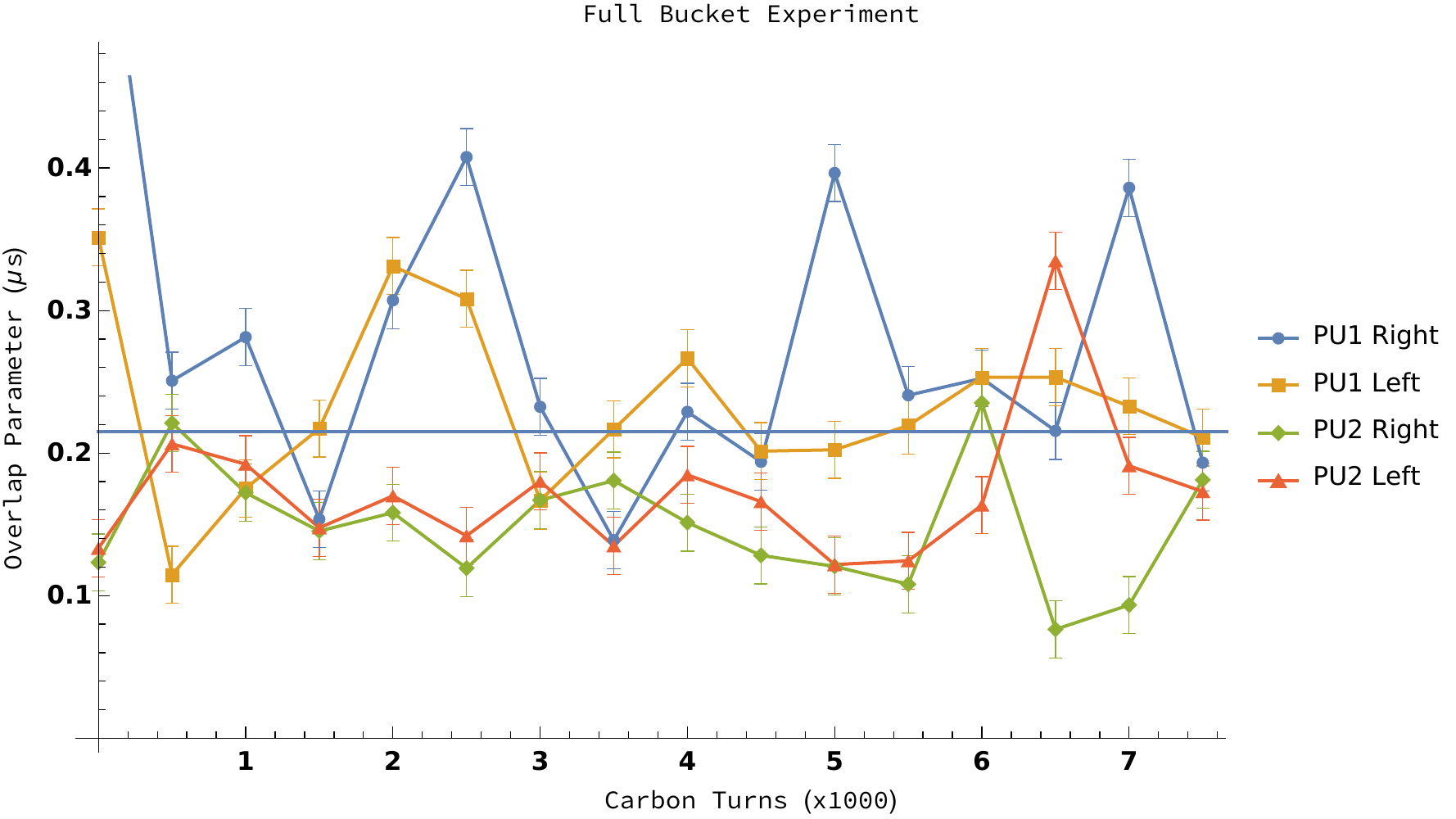}
\caption{These plots show the evolution of the overlap parameter using the constrained fit first for the missing bunch, then for the full bucket system at 7500 carbon turns. The horizontal line shows the predicted overlap parameter.} 
\label{fig_7500_miss_54}
\end{figure}

Interestingly, the bunches do not move symmetrically around the expected value. PU1's data shows a larger value than PU2. This implies that the timing is incorrect, and that the nitrogen bunch is entering the merger region late. The collision is happening, but is not centered in the MR. Plots of both the missing bunch and full bucket measurements in the machine are shown in Fig \ref{fig_7500_miss_54}.

Again we tried to keep the signals symmetric, and after a great deal of scanning phases and frequencies Fig \ref{fig_7500_miss_54} shows the best pattern we could get. The smaller overlap parameter makes these measurements difficult, and using the offset data we got from Fig \ref{fig_950_54_offset} would likely have given us timing that wasn't shifted slightly like we have in this system. This type of analysis should provide a more accurate way to synchronize the bunches.

\section{Potential Interactions}
\label{interactions}

For this experiment we did not expect to measure a direct beam-beam interaction due to the low current and the limited number of diagnostics. Conventional, high energy, beam-beam systems are head-on instead of co-moving, reducing their effects to an impulse, and the beams are usually relativistic enough that the fields can be considered purely transverse. Neither of these is the case in DESIREE. Since the energies, and thus velocities are low enough, there will be longitudinal effects on each bunch.  Two effects can be seen; one is a change to the distribution over repeated collisions, and another is a change in the velocity of the bunches as they move through each other.

\begin{figure}[H]
\centering
\includegraphics[width=0.49\linewidth]{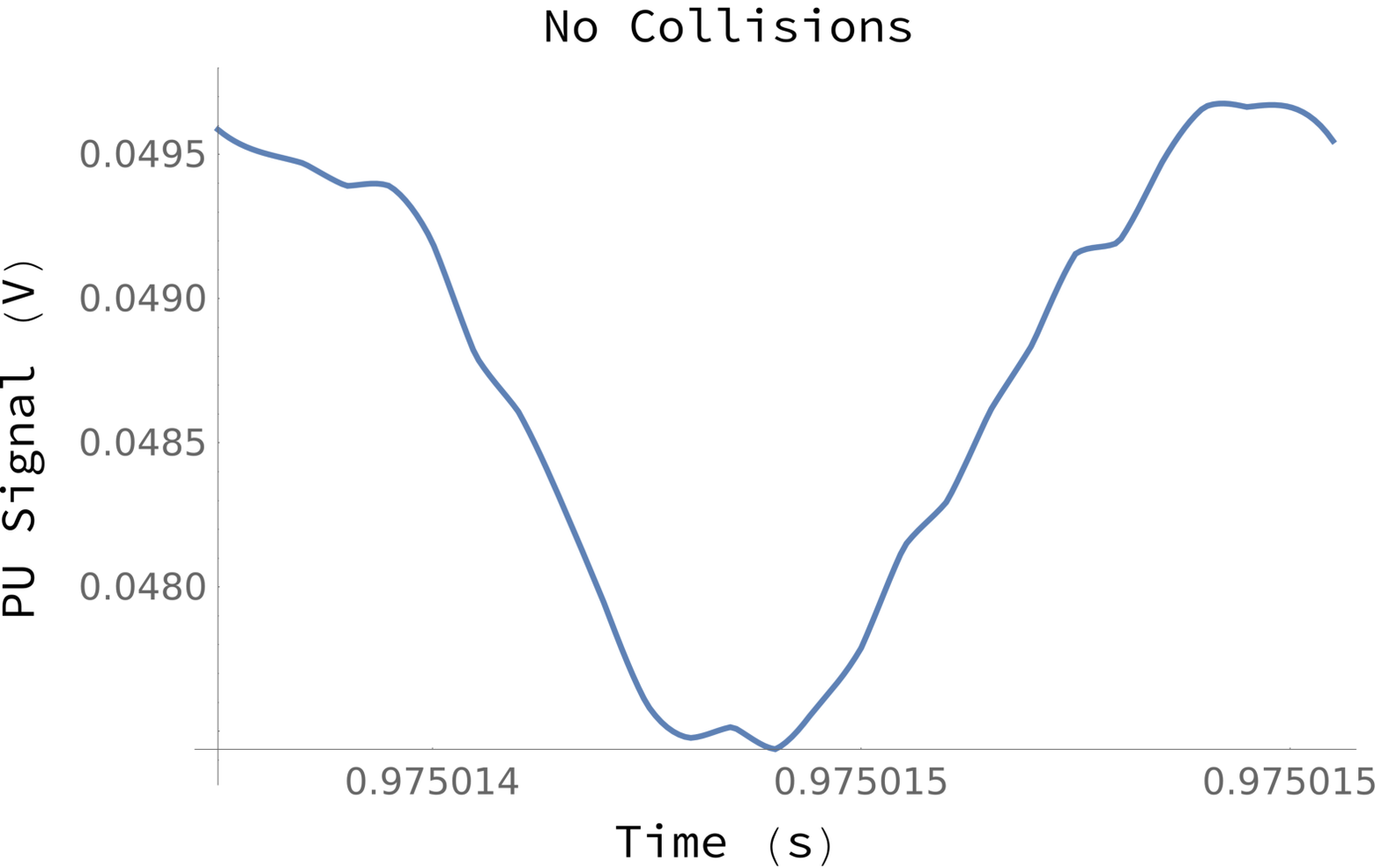}
\includegraphics[width=0.49\linewidth]{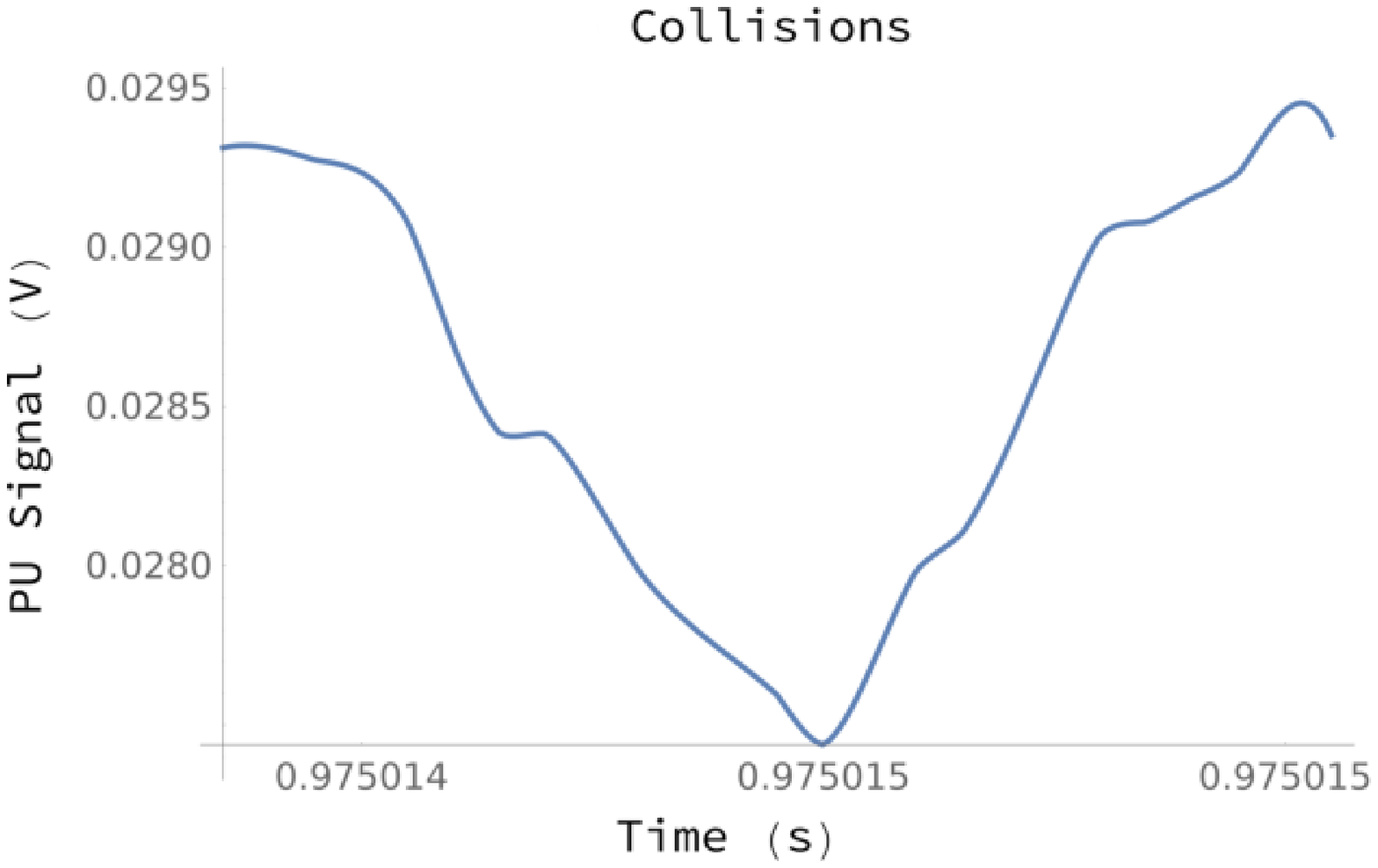}
\caption{Plots of the pickup signal for the nitrogen beam. The first plot has only a nitrogen beam in the machine, the second is the signal of an uncollided bunch in the missing bunch experiment, both after $37500$ carbon turns.} 
\label{fig_bunches}
\end{figure}

An example of the change in the structure of the bunches over multiple turns is shown in Fig \ref{fig_bunches}. A future multi-particle simulation of these interactions should give us a better understanding both of the evolution of these bunches as well as how we may be able to measure these interactions, providing an opportunity for future research.

To find the time of flight difference between a collided and uncollided bunch, we need only solve the system of differential equations shown below to get a first order guess of how the timing will be altered by the interactions of the bunches.
\begin{eqnarray}
Ni''(t) &=& \frac{e^2 N_C}{4 \pi \epsilon_0 A m}\frac{Ni(t)-Ca(t)}{((Ni(t)-Ca(t))^2+\delta^2)^{\frac{3}{2}}}, \\
Ca''(t) &=& \frac{e^2 N_N}{4 \pi \epsilon_0 A m}\frac{Ca(t)-Ni(t)}{((Ni(t)-Ca(t))^2+\delta^2)^{\frac{3}{2}}},
\end{eqnarray}
where $Ni(t)$ and $Ca(t)$ are the positions of the nitrogen and carbon bunch centroids respectively, with $N_N$ and $N_C$ the number of nitrogen and carbon atoms. The term $\epsilon_0$ is the electric permittivity, A is the atomic mass number, and m is the atomic mass, and $\delta$ is a value of $10^{-7}$ to prevent singularities. In the experiment the number of particles per bunch were about 45000 for Carbon and 35000 for Nitrogen. This gives an offset less than $5$ ns. With a higher current and a more accurate oscilloscope, we expect a shift between the uncollided and collided beams of $5$ to $10$ ns which would be measurable.

\section{Discussion, Conclusions, and Future Work}
\label{discussion}

We have described a recent demonstration of a gear-changing system for the first time in an operating machine using DESIREE. The predicted overlap parameter was observed in four on three and five on four setups. There is also evidence suggesting the system has beam-beam interactions occurring, providing an opportunity for future study. 

We learned how to perform collider experiments with beams moving in the same direction. During the experiments,  we checked the timing of collisions by measuring the distances between the positive and negative peaks in the colliding system. Looking back it would have been better for us to use the uncollided signals in the missing bunch experiment to fine tune the timing. These types of lessons are important when controlling a system that hasn't been created before.

While the timing in the five on four system exhibited some error, it did yield useful information about the fitting methods that we used to measure the four on three system. Limited diagnostics are one of the limitations of this experiment, which can be remedied with more dedicated instrumentation, and a more sensitive oscilloscope for future experiments.

We are not sure how much higher we can push the number of bunches using DESIREE. For an $n(n-1)$ system the overlap parameter gets smaller with increasing $n$ to the point that it would become difficult to measure. Measuring the uncollided bunches in the missing bunch experiment will help, but it would require a more complicated pattern for the uncollided system. While the RF system is capable of harmonic numbers as high as 100, the chopper is thermally limited to an unknown degree. Another possibility is a five on three system which would increase the maximum $n$ since the velocity difference, and the overlap parameter, would be larger. We also observed synchrotron motion in the bunch lengths which couldform the basis for future measurements.

While there is potential for ongoing experiments with gear-changing in DESIREE, a better understanding of the beam-beam effect requires an experiment in a machine such as RHIC, or a dedicated facility. A machine with a similar footprint to DESIREE but using electrons, and having dedicated diagnostics would not only be able to perform gear-changing, but would have a beam-beam effect sufficient to reach the predicted instability threshold calculated in \cite{PhysRevSTAB.17.041001}.

\section{Acknowledgements}
The authors would like to thank both Andrew Hutton and Todd Satogata for many useful conversations, and a great deal of advice with this. This work was performed at the Swedish National Infrastructure, DESIREE (Swedish Research Council contract No.2017-00621). This material is based upon work supported by the U.S. Department of Energy, Office of Science, Office of Nuclear Physics under contract DE-AC05-06OR23177. 

\bibliography{bibliographyDODGE}

\end{document}